\documentstyle [manuscript,aps,epsfig]{revtex}
\catcode`\@=11

\draft
\newcommand{\be}{\begin{equation}}
\newcommand{\ee}{\end{equation}}
\begin{document}
\title{Zero-variance principle for Monte Carlo algorithms}

\vspace{2cm}

\author{Roland Assaraf and Michel Caffarel}

\vspace{3cm}

\address{
CNRS-Laboratoire de Chimie Th\'eorique
Tour 22-23, Universit\'{e} Pierre et Marie Curie
4 place Jussieu \\ 75252 Paris Cedex 05; France
e-mail: ra@lct.jusssieu.fr and mc@lct.jusssieu.fr}

\vspace{3cm}

\address{\mbox{ }}
\address{\parbox{14cm}{\rm \mbox{ }\mbox{ }
We present a general approach to greatly increase
at little cost the efficiency of Monte Carlo algorithms.
To each observable to be computed we associate a renormalized observable 
(improved estimator) having the same average but a different variance. By writing 
down the zero-variance condition a fundamental 
equation determining the optimal choice for the renormalized 
observable is derived (zero-variance principle for each observable separately). 
We show, with several examples including classical and quantum Monte Carlo 
calculations, that the method can be very powerful.
}}
\address{\mbox{ }}
\address{\parbox{14cm}{\rm PACS No:02.70.Lq,05.10.Ln,05.50.+q,75.40.Mg}}
\maketitle
\newpage

\makeatletter
\global\@specialpagefalse
\makeatother

Since the pioneering work of Metropolis {\it et al.}\cite{Metro} 
Monte Carlo methods have been widely used in many areas of natural sciences. 
At the root of Monte Carlo methods lies a very efficient stochastic method for calculating 
many-dimensional integrals (or sums) written under the general form
\begin{equation}
<O> \equiv  \frac{ \int_S dx \pi(x) O(x)} {\int_S dx \pi(x)}
\label{defO}
\end{equation}
where $O(x)$ is some arbitrary observable (real-valued function) defined on the 
configuration space $S$ (continuous or discrete) and $\pi(x)$ some probability distribution.
In Monte Carlo methods the integrals are evaluated using a large but finite 
set of configurations $\{ x^{(i)}\}_{i=1,N} $distributed according to $\pi$ and generated 
by a step-by-step stochastic procedure (Markov chain)
\begin{equation}
<O> = \frac{1}{N} \sum_{i=1}^{N} O[x^{(i)}] + \delta O,
\label{defOsim}
\end{equation}
where $\delta O$ is the statistical error associated with the finite statistics. 
For a large enough number $N$ of Monte Carlo steps,
standard statistical arguments lead to the following expression of the error
\begin{equation}
\delta O = K \frac{\sigma(O)}{\sqrt{N}}
\label{delta}
\end{equation}
where $K$ is some positive constant proportional to
the amount of correlation between configurations  and $\sigma(O)$ a measure of 
the fluctuations of the observable
\begin{equation}
\sigma(O) \equiv \sqrt{ <O^2>-{<O>}^2 }.
\label{sigma}
\end{equation}

In this Letter it is shown that by introducing a suitably renormalized observable
${\tilde O}(x)$ the statistical error can 
be drastically reduced and even suppressed, thus defining a zero-variance principle 
for the Monte Carlo calculation of observables.
To realize this, a trial operator $H$ and a trial function 
$\psi(x)$ are introduced (a trial matrix and a trial vector in the discrete case).
The operator $H$ is supposed to be Hermitian (in all practical applications, real symmetric) 
and is chosen such that
\begin{equation}
\int dy H(x,y)\sqrt{\pi(y)} = 0.
\label{Hdef}
\end{equation}
On the other hand, the trial function $\psi(x)$ is a rather arbitrary function 
which is simply supposed to be integrable. Now, the renormalized observable
${\tilde O}(x)$ associated with the observable $O(x)$ is defined as follows
\begin{equation}
{\tilde O}(x) = O(x) + \frac{ \int dy H(x,y)\psi(y)} {\sqrt{\pi(x)}}.
\label{Otilde}
\end{equation}
As a direct consequence of Eq. (\ref{defO}) and of the very
definition of the Hermitian operator $H$, Eq. (\ref{Hdef}), we have the important property
\begin{equation}
<{\tilde O}> = <O>.
\label{i}
\end{equation}
In other words, both quantities $O(x)$ and ${\tilde O}(x)$ can be used as estimators
of the desired average. 
However, the statistical errors, which are controlled by $\sigma(O)$ and $\sigma({\tilde O})$,
can be very different. The optimal choice for $(H,\psi)$ is obtained by imposing 
the renormalized function to be constant and equal to the exact average.
This leads to the following fundamental equation
\begin{equation}
\int dy H(x,y) \psi(y)= -[O(x)-<O>]\sqrt{\pi(x)}  \Leftrightarrow \sigma({\tilde O})=0
\label{eqfond}
\end{equation}

At this point it should be emphasized that the idea of using renormalized estimators 
for reducing the variance is not new. A number of applications have been performed using 
various ``improved'' estimators having a lower variance (See, e.g.,\cite{improv},\cite{improv2}). 
The basic idea is to construct new estimators
by integrating out some intermediate degrees of freedom and, therefore, 
removing the corresponding source of fluctuations.
However, to the best of our knowledge, no general and systematic approach based on a 
zero-variance principle and valid for any type of Monte Carlo methods has been proposed so far.

In this work the following strategy is proposed.
First, a Hermitian operator $H$ verifying (\ref{Hdef}) is chosen. Second, some approximate 
solution of Eq.(\ref{eqfond}) is searched for. The various parameters entering $\psi$ are 
then optimized by minimizing the fluctuations of the renormalized observable over a finite 
set of points distributed according to $\pi$ and obtained from a short Monte Carlo calculation. 
Finally, a standard much longer Monte Carlo simulation 
is performed using ${\tilde O}(x)$ instead of $O(x)$ as estimator.\\
{\it Choice of $H$.}
Clearly, a large variety of choices are possible for the trial operator $H$. For Monte Carlo
algorithms satisfying the detailed balance condition (in practice, the vast majority of 
MC schemes) a very natural choice is at our disposal. Denoting $ p(x \rightarrow y)$ the 
transition probability distribution defining the Monte Carlo dynamics, the detailed balance 
condition is written as
$\pi(x) p(x \rightarrow y)= \pi(y) p(y \rightarrow x)$
for all pairs (x,y) in configuration space.
A most natural operator to consider is 
\begin{equation}
\label{hdef}
H(x,y)= \sqrt{ \frac{\pi(x)}{\pi(y)} } [ p(x \rightarrow y) - \delta(x-y) ].
\label{hxy}
\end{equation}
From the detailed balance condition it follows that the operator $H$ is symmetric, 
$H(x,y)=H(y,x)$. The fundamental 
property (\ref{Hdef}) is verified since the sum-over-final-states for a transition probability 
is equal to one. For continuous systems Schroedinger-type Hamiltonians can also 
be considered
\begin{equation}
H = -\frac {1}{2} \sum_{i=1}^{d} \frac{{\partial}^2}{\partial {x_i}^2} + V(x)
\label{Hsdef}
\end{equation}
where $V(x)$ is some local potential constructed to fulfill condition (\ref{Hdef}):
\begin{equation}
V(x)=  \frac {1}{2 \sqrt{\pi(x)}} 
\sum_{i=1}^{d} \frac{{\partial}^2 \sqrt{\pi(x)} }{\partial {x_i}^2},
\label{Hvdef}
\end{equation}
where $d$ is the number of degrees of freedom. Note that in Eq. (\ref{Hsdef}) 
$H$ is written using the standard quantum-mechanical notation for a {\sl local} Hamiltonian 
in the $x$-space realization. 
\\
{\it Choice of $\psi$.}
Once the operator $H$ has been chosen, the optimal choice for $\psi$ is the 
exact solution of the fundamental equation.
Of course, in practice only approximate solutions are available. 
What particular form to choose for $\psi$ is 
very dependent on the problem at hand, 
on the type of observables considered, and also on the form chosen for the trial operator $H$. 
However, a most important point to be stressed is that the global normalization factor 
associated with $\psi$ is a pertinent parameter of the trial function.
Minimizing the fluctuations of the 
renormalized function $\sigma({\tilde O})$ with respect to it, we get
\begin{equation}
{\sigma({\tilde O})}^2=
{\sigma(O)}^2
- \frac{  { < \frac{ O(x) \int dy H(x,y)\psi(y)} {\sqrt{\pi(x)}} > }^2  }
          { < \biggl({  \frac{ \int dy H(x,y)\psi(y)} {\sqrt{\pi(x)}}   }\biggr)^2 > }.
\label{sigma2}
\end{equation}
The correction to ${\sigma(O)}^2$ being negative we obtain the important result that,
whatever the choice made for the trial function (even the most unphysical one!), 
the optimization of the multiplicative factor always leads to a reduction 
of the statistical error.

Our first application concerns the Monte Carlo calculation of the internal energy of the standard 
$2D$-Ising model at various temperatures and linear sizes $L=5,10,20$, and $25$. 
The observable considered is
the energy function given by $E(S)= -\sum_{<i,j>} S_i S_j$ 
(coupling constant $J=1$, sum limited to nearest neighbors, and periodic boundary conditions).
The probability distribution is
$\pi(S)= \exp[-\beta E(S)]$ with $\beta= 1/k_B T$. 
Here, $S \equiv (S_1,\cdots,S_N)$ with $S_i=\pm 1$, and $N=L \times L$ is the total number of spins. 
Simulations have been performed using a Swendsen-Wang type 
algorithm \cite{swendsen}(non-local updates of clusters of spins).
To construct the trial operator $H$ we have chosen to 
use the transition probability distribution of Monte Carlo algorithms
with local updates (``Heat-Bath''-type algorithms). The probability of 
flipping the spin $S_i= \pm 1$ at site $i$ is given by
\begin{equation}
p(S_i \rightarrow \epsilon S_i) = \frac{ e^{\beta \epsilon S_i {\tilde S_i} }}
  { e^{\beta S_i {\tilde S_i} }  + e^{-\beta  S_i {\tilde S_i} }     }
\label{pijlocal}
\end{equation}
where $\epsilon$=1 (no flip) or -1 (flip), and ${\tilde S_i}$ is the 
sum of neighboring spin values. 
With this choice and using Eq.(\ref{hxy}) the fundamental equation (\ref{eqfond})
can be rewritten under the form 
$$
\sum_{i=1}^N p(S \rightarrow T_iS)
 [Q(S)-Q(T_iS)]= E(S)-<E>
$$
\begin{equation}
\psi(S) = Q(S) \sqrt{\pi(S)}
\label{eqex}
\end{equation}
where the application $T_i$ ($i=1,\cdots,N$) describes a flip at site $i$ and is defined by 
$T_i(S_1,\cdots,S_i,\cdots,S_N)=(S_1,\cdots,-S_i,\cdots,S_N)$.
At $\beta= 0$ ($T=\infty$) the transition probability distribution becomes constant and 
the exact solution is easily found to be $Q(S)= E(S)/2$. 
For finite temperatures some approximate solution has to be found. Here we introduce
for $Q(S)$ a polynomial expansion up to the fourth-order in the variables 
$X=\sum_{i=1}^N S_i$ (magnetization) and $Y=\sum_{i=1}^N g( S_i {\tilde S_i})$ 
(``generalized energy'', the usual energy being recovered for $g(x)=x$). Precisely,
we have chosen the form $Q(S)=e^{-Z} \sum_{n+m \le 4} c_{nm} X^n Y^m$ where 
$Z=\sum_{i=1}^N h( S_i {\tilde S_i})$. The set of variational parameters of 
$\psi$ consists of all coefficients $c_{nm}$ of the polynomial plus the ten possible values of 
functions $g$ and $h$. All coefficients have been optimized by minimizing  
the fluctuations of the renormalized energy $\sigma({\tilde E})$ defined by (\ref{sigma})
and calculated from $2000$ to $5000$ different spin configurations $S^{(i)}$ drawn 
according to $\pi$. Finally, the last step consists in performing a 
long Monte Carlo simulation to compute accurately the various quantities. The number of 
clusters built varies from $10^6$ (for the larger size) to $2. 10^8$ (for the smaller size).
Results are presented in Table \ref{ising2D}. Three different temperatures have been 
considered. $T=3$ corresponds to the low-temperature regime, $T=T_c=4/\ln(\sqrt{2}+1)$ is 
the critical temperature for the infinite lattice, and $T=8$ is in the high-temperature regime
of the model. At $T=3$, our representation is extremely good whatever the size of the lattice 
considered. The variance associated with the renormalized energy is drastically reduced 
with respect to the bare value and the gain in computational 
effort can be as great as $\sim 360$. Here, the gain in computational effort is defined as
the ratio of the squared statistical errors,${(\delta {\tilde E}/{\delta E})}^2$. In other 
words, according to Eq.(\ref{delta}) it represents the factor by which it would be necessary to increase 
the number of Monte Carlo steps in the standard approach to get the same accuracy.
Note that for $L=5$ our Monte Carlo value coincides with the 
exact one (computed by exact numeration of the $2^N$ configurations) with an accuracy of less 
than $10^{-6}$. Note also that our MC values converge as the size is increased to the exact
infinite-lattice value as given by the Onsager solution \cite{onsager}.
At $T=8$ (high-temperature regime) our representation is less good but still very
satisfactory. As a function of the size, the gain in computational effort converges and 
a value of about 20 is gotten. At the infinite-lattice critical value the results are less 
spectaculary but still of interest. A converged value of about $3$ for the gain 
in efficiency is obtained. At this temperature the correlation length for the 
spin variables diverges and more accurate representations for the solution 
of Eq.(\ref{eqex}) are needed. 
Starting from our basic equation built from a transition probability corresponding to
{\it local} moves we need to resort to approximate solutions which contain in some way 
the collective spin excitations. Alternatively, we can change our fundamental equation 
by resorting to a non-local transition probability density and, then, to a new operator $H$. 
A natural choice is of course the transition probability
of the Swendsen-Wang algorithm  used here to generate 
configurations. Preliminary calculations show that statistical fluctuations are indeed
strongly decreased. However, to sum up analytically  all 
contributions corresponding to the different Swendsen-Wang clusters (action of $H$ on $\psi$) 
is very time-consuming and the advantages of the method can be lost. Some approximate scheme 
is clearly called for; this is let for future development. 
Finally, a last important point is that the gain in computational effort is found 
to be systematically greater (by about $50\%$) than the corresponding ratio of variances. 
This result is a direct consequence of the fact that the integrated autocorrelation 
time known to control the amount of correlation between successive 
measurements (see, e.g., \cite{improv2}) has been decreased when passing from the 
bare observable to the renormalized one. Note that a similar behavior has also been 
obtained in applications based on improved estimators \cite{improv},\cite{improv2}. 
Without entering into the details, it can be shown that this result is directly related 
to the fact that the fluctuations of the renormalized observable are much smaller than in 
the bare case.

The second application illustrates the method in the case of a continuous configuration 
space (calculation of multi-dimensional integrals). We have calculated a
mean energy as it appears in the so-called Variational Monte Carlo (VMC) methods
\cite{vanlinden}. Starting from a quantum Hamiltonian $H_Q$ 
(to be distinguished from our trial operator $H$) and a known trial-wave function 
$\psi_T$ our purpose is to compute the variational energy $E_v$ associated with $\psi_T$.
$E_v$ can be easily rewritten as an average over 
the probability distribution ${\psi_T}^2$, $E_v=<E_L>$
where $E_L \equiv H_Q\psi_T/\psi_T$ is called the local energy.
Here, we consider the case of the Helium atom described by the Hamiltonian 
$H_Q= -1/2 ({\vec{\nabla_1}}^2 + {\vec{\nabla_2}}^2) -2/r_1 -2/r_2 +1/r_{12}$ (atomic units) 
with usual notations. As trial wave function a standard form has been chosen \cite{Mosko}
\be
\psi_T (\vec{r}_1,\vec{r}_2)
=\exp[ \frac{a r_{12}}{1+b r_{12}}-c (r_1+r_2) ] 1s(r_1) 1s(r_2)
\ee
where $1s(r)$ is the Hartree-Fock orbital as given by Clementi and Roetti \cite{Clementi}
and the variational parameters have been chosen to be $a=0.5,b=0.522$, and $c=0.0706$.
As already remarked a natural choice for the trial Hamiltonian $H$ 
is a Schroedinger operator admitting $\psi_T$ as ground-state, Eqs.(\ref{Hsdef},\ref{Hvdef}). 
Regarding $\psi$ we have chosen a form similar to the trial wavefunction 
multiplied by some function of the potential energy.
Configurations are generated using a standard Metropolis algorithm with 
local moves constructed using a Langevin equation \cite{Mosko}.
Results are presented in Table \ref{He}. It is seen that the introduction of the 
renormalized local energy increases the 
efficiency of the Monte Carlo calculation by about one order of magnitude. 

In the last application it is shown that the method can even be used in 
exact (zero-temperature) quantum Monte Carlo (QMC) calculations. In QMC a combination 
of diffusion and branching process 
is used to construct a stationary density proportional to $\psi_T \psi_0$ where $\psi_0$ is the 
exact unknown ground-state wavefunction. By averaging the local energy over this distribution 
an estimate of the exact energy $E_0$ is obtained \cite{vanlinden}.
Although the analytical form of the stationary 
density is no longer known, a renormalized function
whose average is identical to that of the bare local energy can still be defined,
${\tilde E_L} = E_L + {(H-E_0)\psi}/{\psi_T}$,
where $H$ admits $\psi_T$ as eigenvector, $H\psi_T=0$.
Calculations have been done using the exact
Green's function Monte Carlo of Ceperley and Alder \cite{ceperley}. 
Results are presented in Table \ref{He}. 
They are of a quality similar to that obtained in the variational case.
About one order of magnitude in computer time has been gained.

To conclude we have presented a simple and powerful method to greatly increase at little cost 
the efficiency of Monte Carlo calculations. The examples presented have been chosen to 
illustrate the great versatility of the method (discrete and continuous configuration spaces, classical 
or quantum Monte Carlo, local or non-local Monte Carlo updates). 
Although our examples have only been concerned with total energies, let us emphasize that
the zero-variance principle is valid for any type of observable including
important quantities such as local properties other than energy, 
differences of energies, spatial correlation functions, etc...

\newpage

\begin{table}[h]
\caption{Internal Energies for the 2D-Ising model at different temperatures. $N$ number 
of sites. Statistical uncertainties on the last digit are indicated in parentheses.}
\begin{tabular}{lccccc}
\hline
\multicolumn{1}{l}{Size}&
\multicolumn{1}{c}{$ 5 \times 5$}&
\multicolumn{1}{c}{$10 \times 10$}&
\multicolumn{1}{c}{$20 \times 20$}&
\multicolumn{1}{c}{$25 \times 25$}&
\multicolumn{1}{c}{$\infty \times \infty$}\\
\hline
\hline
$T=3$       &   &   &   &   &  \\
\hline
\hline
${\sigma(E)}^2/N$& 1.789(1) & 1.777(2)      & 1.78(1)       & 1.79(1)      & \\
${\sigma({\tilde E})}^2/N$& 0.0125(4)& 0.0061(1)  & 0.0060(2) & 0.0061(2)    & \\
Ratio of variances         & $\sim 143$    & $\sim 291$    & $\sim 297$    & $\sim 293$   & \\
$<E/N>$                    & -3.902044(31) & -3.902200(55)& -3.90217(21.4)& -3.90242(29) & \\
$<{\tilde E}/N>$           & -3.902020(2.4)& -3.902229(3)& -3.90225(1.2) & -3.90222(1.5)&  \\
Gain in computational effort$^a$& $\sim 167$& $\sim 336 $ & $\sim 318$    &  $\sim 360$  &  \\
$<E/N>$ Exact value         & -3.9020214... &               &               &  & -3.9022331...$^b$\\
\hline
\hline
$T=T_c=4.53837...$       &   &   &   &   &  \\
\hline
\hline
${\sigma(E)}^2/N$            &  18.581(4)  & 25.97(3)   & 33.1(2)   & 35.3(2)     & \\
${\sigma({\tilde E})}^2/N$   &   0.215(2)  &  4.85(1)   & 16.5(1)   & 16.9(2)     & \\
Ratio of variances         & $\sim 86$   & $\sim 5.4$ &$\sim 2.0$ & $\sim 2.1$    & \\
$<E/N>$                    &-3.07334(13) &-2.95214(33)&-2.8902(12)&-2.8800(14)  &  \\
$<\tilde{E}/N>$            &-3.07345(1.3)&-2.95236(13)&-2.8908(7) &-2.8788(8)  &  \\
Gain in computational effort$^a$& $\sim 100$  & $\sim 6.8$  & $\sim 3.$ & $\sim 3.1$  &  \\
$<E/N>$ Exact value        &-3.0734396...&            &           &             &-2.8284271...$^b$\\
\hline
\hline
$T=8$       &   &   &   &   &  \\
\hline
\hline
${\sigma(E)}^2/N$            &  13.17(1)    & 10.96       & 11.1(2)   & 10.9(3)      & \\
${\sigma({\tilde E})}^2/N$   &   0.041      &  0.455      &  0.8(1)   &  0.9(1)      & \\
Ratio of variances         &  $\sim 321.2$&  $\sim 24.$ &$\sim 13.9$& $ \sim 12$   & \\
$<E/N>$                    & -1.16440(33) & -1.11556(48) & -1.1156(20)&-1.1165(25.6)& \\
$<\tilde{E}/N>$            & -1.16348(1.6)& -1.11502(8.2)& -1.1145(4.4)&-1.1145(5.6)&\\
Gain in computational effort$^a$&  $\sim 425$  &  $\sim$ 34  & $\sim 20.7$& $\sim 20.9$ & \\
$<E/N>$ Exact value        & -1.1634926...&             &            &              &-1.1145444...$^b$\\
\hline
\end{tabular}
\raggedright
\small{$^a$ See text for definition.}\\
\small{$^b$ Ref.\cite{onsager}.}\\
\label{ising2D}
\end{table}
\begin{table}[h]
\caption{Energy of the Helium atom. All quantities are given in atomic units.
Statistical uncertainties on the last digit are indicated in parentheses.}
\begin{tabular}{lc}
Variational Monte Carlo    &              \\
\hline
\hline
${\sigma(E_L)}^2$          & 0.0409(2)    \\
${\sigma({\tilde E_L})}^2$ & 0.00688      \\
Ratio of variances         & $\sim 5.9$   \\
$<E_L>$                    & -2.89671(4.8)\\
$<\tilde{E_L}>$              & -2.89674(1.6)\\
Gain in computational effort$^a$& $\sim 9$     \\
\hline
\hline
Exact Green's Function Monte Carlo$^b$    & \\
\hline
\hline
${\sigma(E_L)}^2$          & 0.0411(9)  \\
${\sigma({\tilde E_L})}^2$ & 0.00855(8)  \\
Ratio of variances         &  $\sim 4.9$ \\
$<E_L>$                    & -2.903745(99)\\
$<\tilde{E_L}>$              & -2.903734(33)\\
Gain in computational effort$^a$& $\sim 9$  \\
Exact energy               &${-2.903724377...}^c$\\
\end{tabular}
\raggedright
\small{
$^a$ See text for definition.\\
$^b$Reference \cite{ceperley}\\
$^c$Reference \cite{Heex}\\
      }
\label{He}
\end{table}

\end{document}